\documentclass[preprint]{vgtc}               

\title{Uncertainty-Aware Jacobi Set Computation} 

\author{%
	\authororcid{Daniel Klötzl}{0000-0002-4222-3320}\thanks{University of Stuttgart - firstname.lastname@visus.uni-stuttgart.de}
	\and
	\authororcid{Daniel Weiskopf}{0000-0003-1174-1026}\footnotemark[1]
}

\abstract{
	We present an uncertainty-aware Jacobi set computation method. In general, Jacobi sets are topological descriptors that capture the gradient alignments of two scalar fields, as, e.g., used for multi-field visualization. We adopt and reformulate an existing computational approach that relies on an edge-based identification of Jacobi set edges on a given triangulation. Our extension to uncertainty visualization builds upon a versatile, spatially coherent uncertainty model for pairs of scalar fields based on multivariate normal distributions. We propagate the uncertainty analytically, thereby lifting the original Jacobi set computation to uncertain inputs. Furthermore, we present an overlay of visual mappings specifically designed to show the Jacobi sets along with different facets of uncertainty information. Both the uncertainty model and uncertainty-aware method are validated against a Monte Carlo approach on an analytic dataset and applied to two use cases from fluid dynamics and weather ensembles.
}


\teaser{
	\centering
	\includegraphics[width=\linewidth]{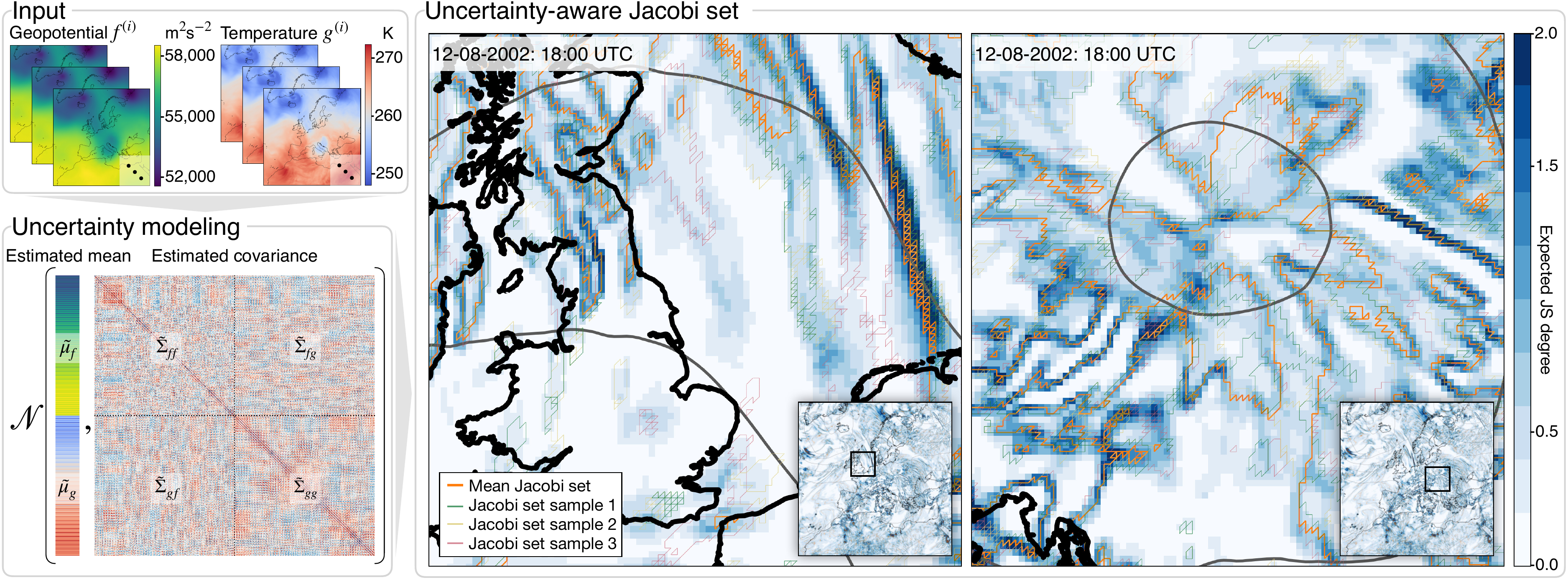}
	\caption{Uncertainty-aware Jacobi set computation for ensemble data modeled as a multivariate normal distribution. 
		(Left, top)~The CERRA-EDA weather ensemble~\cite{Ridal2024} consists of 10 members for each of the scalar fields geopotential (in m$^\text{2}$\,s$^\text{-2}$) and temperature (in K) at 500\,hPa over Europe, 12~August 2002. 
		(Left, bottom) The members are aggregated into the stacked mean fields and covariance matrix, shown normalized to correlations with a diverging blue--red colormap; the off-diagonal blocks hold the cross-field covariance $\tilde\Sigma_{fg}$ between geopotential and temperature.
		(Center and right) The uncertainty-aware Jacobi set visualization for two zoomed-in areas consists of the Jacobi set of the mean fields (in orange), the expected Jacobi set degree (via the heatmap with bluish color map), and three independently and randomly drawn Jacobi set samples (green, sand, and rose), which indicate the realization-level variability.
		Coastlines (black) and mean geopotential isolines~(gray) provide geographic and meteorological context.
	}
	\label{fig:teaser}
}

\usepackage{amsmath, amssymb}
\usepackage{mathptmx}   
\usepackage{algorithm}
\usepackage{algpseudocode}
\usepackage{tikz}
\usetikzlibrary{arrows.meta,calc}

\definecolor{jsorange}{RGB}{230,120,20}
\definecolor{cwblue}{RGB}{59,76,192}   
\definecolor{cwred}{RGB}{180,4,38}     

\newcommand{\R}{\mathbb{R}}
\newcommand{\etal}{et~al.}
\newcommand{\E}{\mathop{}\operatorname{E}}
\newcommand{\Var}{\mathop{}\operatorname{Var}}
\newcommand{\Cov}{\mathop{}\operatorname{Cov}}
\newcommand{\tr}{\mathop{}\operatorname{tr}}

\usepackage{silence}
\AtEndPreamble{\hypersetup{pdfauthor={Daniel Kl\"otzl, Daniel Weiskopf}}}

\begin{document}
	\firstsection{Introduction}
	
	\maketitle
	
	Phenomena in fluid dynamics, weather forecasting, oceanography, or climate science are often described by two or more scalar fields on a common domain. 
	The Jacobi set, originally introduced by Edelsbrunner and Harer~\cite{Edelsbrunner2002}, is a topological descriptor consisting of all points at which the gradients of two scalar fields are aligned.
	In practice, these fields often stem from simulation data with uncertain inputs or forecasts and are given as ensemble simulations or with known model variability.
	The original Jacobi set, as many topological descriptors, is very sensitive to uncertainty because it is defined through piecewise linear derivatives such that small perturbations can change the extracted structures.
	Given uncertain scalar fields, one could compute the Jacobi set from the mean fields, which would result in a clean but misleading visual representation.
	An honest analysis would therefore propagate the input uncertainty through the Jacobi set computation, letting us visualize the position of the Jacobi set together with the certainty in this position.
	
	An appropriate uncertainty description serves as the foundation for such uncertainty visualization.  
	Existing approaches to uncertainty in scalar or vector field data often use uncorrelated uncertainty models, i.e., only the mean and variance of the underlying data, while neglecting spatial coherence.
	These models inherently produce inconsistent field realizations, e.g., visible as zig-zag streamlines~\cite{Otto2010, 2015MathiasOttoDiss}. 
	The importance of considering spatial correlation for uncertain field visualization has been demonstrated, e.g., for isocontours by Pfaffelmoser~\etal~\cite{Pfaffelmoser2011, Pfaffelmoser2012}, and using multivariate normal distributions with full covariance matrices leads to coherent realizations, as applied to time series by Krake~\etal~\cite{Krake2025}.
	However, regardless of the uncertainty model, no uncertainty-aware computation of Jacobi sets exists so far, and the only strategy is Monte Carlo (MC) sampling, which requires many Jacobi set extractions.
	
	We close this gap with an analytic \emph{uncertainty-aware Jacobi set} computation, structured in three steps.
	First, we introduce a flexible uncertainty model that describes a pair of scalar fields as a joint multivariate normal distribution over the vertices of a triangulation.
	This model can be estimated from ensemble data or constructed via kernel functions from expert knowledge, and it also covers correlations between the two scalar fields.
	Second, the edge-based identification of Jacobi set edges by Edelsbrunner and Harer~\cite{Edelsbrunner2002} is reformulated by propagating the uncertainty through the piecewise linear computation and lifting the critical edge test to a per-edge Jacobi set crossing probability.
	Third, we visualize the Jacobi set of the mean fields 
	(in short, the mean Jacobi set) and the probabilistic Jacobi set by aggregating the edge probabilities to vertex-based encodings.
	
	\Cref{fig:teaser} illustrates this pipeline for a weather ensemble.
	Ten members of two scalar fields (left) are aggregated into a joint multivariate normal distribution, also covering correlations between the two fields ($\tilde\Sigma_{fg}\neq0$).
	Our analytic propagation turns this model into the uncertainty-aware Jacobi set visualization (right), consisting of the mean Jacobi set (in orange), the heatmap that shows where Jacobi set edges occur with high probability, and individual realizations that indicate the variability of the extracted Jacobi sets.
	
	In the evaluation, we verify the method visually and quantitatively against an MC approach on an analytic dataset.
	Furthermore, two application scenarios are described.
	First, we apply our method to a weather data ensemble, aggregating ten members of the two scalar fields of temperature and geopotential into a single view.
	Second, a user quantifies the uncertainty of a fluid flow simulation via kernel functions and explores how the Jacobi set structure changes for differently modeled uncertainty. 
	
	The main contributions are:
	\begin{itemize}
		\item a multivariate Gaussian uncertainty model for pairs of scalar fields,
		\item an analytic uncertainty-aware Jacobi set computation that lifts the critical edge test of the piecewise linear method to edge-crossing probabilities, and
		\item a qualitative and quantitative evaluation via comparison with MC solutions and case studies.
	\end{itemize}
	
	\section{Related Work}
	Our work connects the computation of Jacobi sets in multi-field topology and the modeling, propagation, and visualization of uncertainty in scalar and vector fields.
	
	\paragraph{Jacobi Sets and Multi-Field Topology.}
	Edelsbrunner and Harer~\cite{Edelsbrunner2002} introduced Jacobi sets together with an edge-based computation method for piecewise linear functions on triangulated \mbox{2-manifolds}.
	They have since been used to relate scalar fields~\cite{EdelsbrunnerHarerNatarajan2004}, detect ridge and valley structures in image data~\cite{NorgardBremer2012}, track burning regions in time-dependent combustion simulations~\cite{BremerBringaDuchaineau2007}, and analyze dependencies between geophysical multi-fields~\cite{Aramonova2017}.
	The restriction of the extracted set to the triangulation edges produces zig-zag artifacts at grid resolution that can obscure the analysis~\cite{BhatiaWangNorgard2015}.
	Klötzl~\etal~\cite{Kloetzl2022} addressed this issue by using local bilinear interpolation and improving the geometry without changing the topology of the piecewise linear method. Overlapping line segments can be removed through a homotopy-equivalent representation~\cite{Kloetzl2022a}.
	Another option is the direct simplification of the Jacobi set~\cite{NagarajNatarajan2011, BhatiaWangNorgard2015} or indirectly by smoothing the fields~\cite{BremerBringaDuchaineau2007, LuoSafWan2009, Raith2024}.
	Meduri~\etal~\cite{Meduri2024} simplified Jacobi sets based on the robustness of critical points to support tracking over time, and Ma~\etal~\cite{Ma2025} computed Jacobi sets from higher-order functional data approximations.
	For other related multi-field descriptors such as the parallel vector operator~\cite{Peikert1999}, the Reeb space~\cite{EdelsbrunnerHarerPatel2008}, or Pareto sets~\cite{Huettenberger2013}, we refer to the overview by Yan~\etal~\cite{Yan2021b}.
	
	\paragraph{Uncertain Scalar and Vector Field Topology.}
	The modeling and propagation of uncertainty is an active research area in the visualization community.
	Sources for uncertainty throughout the visualization pipeline are measurement errors during the acquisition, model or simulation variability often captured by ensemble runs, or subsequent processing steps such as interpolation or other data transformations~\cite{Bonneau2014}.
	While uncertainty is still frequently omitted in practice~\cite{Hullman2020}, new methods have been developed in recent years and are covered in multiple surveys~\cite{Brodlie2012, Kamal2021, Padilla2021, Haegele2022, Weiskopf2022}.
	For topological features of uncertain scalar fields, Pöthkow and Hege~\cite{Pothkow2011} derived probabilistic measures for the positional uncertainty of isocontours under Gaussian models, later generalized to nonparametric models~\cite{Pothkow2013, Athawale2016}.
	Closer to our setting, Athawale~\etal~\cite{Athawale2025} derived, without sampling, the probability that a vertex of an uncertain 2D scalar field forms a local extremum or saddle.
	For vector fields, Otto~\etal~\cite{Otto2010, 2015MathiasOttoDiss} introduced an uncertain vector field topology.
	These approaches primarily model the uncertainty per vertex (by a mean and a variance) without considering any (spatial) correlation.
	In contrast, Pfaffelmoser~\etal\ studied spatial correlation on the positional variability of isocontours~\cite{Pfaffelmoser2011} and visualized global correlation structures in uncertain 2D scalar fields~\cite{Pfaffelmoser2012}, but not for topological multi-field descriptors.
	To the best of our knowledge, no existing approach incorporates the uncertainty of two input fields and quantifies the resulting uncertainty of the Jacobi set.
	
	\paragraph{Gaussian Uncertainty Propagation.}
	More recent publications model uncertainty using multivariate normal distributions and propagate this uncertainty through analysis pipelines instead of relying on per-realization sampling.
	Examples include principal component analysis~\cite{Goertler2020, Zabel2024, Kloetzl2025}, multidimensional scaling~\cite{Haegele2023}, and spectral analysis~\cite{Evers2025}.
	We refer to the UADAPy toolbox~\cite{UADAPy}, which collects several of these techniques.
	For time series, Krake~\etal~\cite{Krake2025} propagated multivariate normal distributions through the linear seasonal-trend decomposition method.
	These works motivated both the modeling of our uncertain scalar fields (via ensembles or kernel functions) and utilizing that Gaussian distributions are closed under affine maps.
	However, none of these propagation approaches target topological descriptors such as Jacobi sets.
	
	\section{Background}
	\label{sec:background}
	This section recaps the definition of Jacobi sets and their piecewise linear (PL) computation by Edelsbrunner and Harer~\cite{Edelsbrunner2002}.
	In addition, probabilistic foundations used for uncertainty modeling and propagation are described, serving as a basis for our new method.
	
	\subsection{Definition of Jacobi Sets}
	\label{sec:js-def}
	In the following, Jacobi sets are defined on 2D manifolds.
	Given two scalar-valued Morse functions $f, g\colon \mathbb{M} \to \mathbb{R}$ on a smooth manifold~$\mathbb{M}$, the Jacobi set is the set of points where their gradients are linearly dependent~\cite{Edelsbrunner2002}:
	\begin{align*}
		\mathbb{J}(f,g) :=\;& \{p \in \mathbb{M} \mid \nabla f(p) + \lambda \nabla g(p) = 0 \text{ or } \notag \\
		& \phantom{\{p \in \mathbb{M} \mid {}}\lambda \nabla f(p) + \nabla g(p) = 0,\ \lambda \in \mathbb{R}\}\,.
	\end{align*}
	For $\mathbb{M} \subset \mathbb{R}^2$, the linear dependency can be formulated using the \emph{gradient alignment value}~\cite{Kloetzl2022}
	\begin{equation*}
		\kappa_p(f, g) := \partial_x f(p)\, \partial_y g(p) - \partial_y f(p)\, \partial_x g(p)\,,
	\end{equation*}
	where $p\in\mathbb{M}$.
	This leads to the equivalent formulation $\mathbb{J}(f, g) = \overline{\{p \in \mathbb{M} \mid \kappa_p(f, g) = 0\}}$ as the closure of the zero set of the gradient alignment field~\cite{Kloetzl2022, Kloetzl2022a}.
	The Jacobi set is symmetric, $\mathbb{J}(f, g) = \mathbb{J}(g, f)$, and generically a smoothly embedded 1D manifold in $\mathbb{M}$~\cite{Edelsbrunner2002}.

	\subsection{Piecewise Linear Jacobi Sets}
	\label{sec:bg-pljs}
	
	The PL Jacobi set computation follows Edelsbrunner and Harer~\cite{Edelsbrunner2002} on a triangulation $K$ of the manifold $\mathbb{M}\subset\R^2$.
	Both the extracted PL Jacobi set and our later derived edge probabilities are defined relative to this triangulation.
	We use a fixed diagonal direction for regular grids and leave the study of the influence of triangle grid resolution and connectivity to future work.
	For an in-depth motivation and derivation, we refer to Edelsbrunner and Harer~\cite{Edelsbrunner2002} and the background section of Klötzl~\etal~\cite{Kloetzl2022}.
	
	Let $f, g\colon K \to \mathbb{R}$ be the piecewise linear extensions at the vertices $\operatorname{Vert}(K)$.
	Within each triangle $T = \langle a, b, v \rangle$, the PL gradients $\nabla f|_T$ and $\nabla g|_T$ are constant.
	Thus, the gradient alignment field is piecewise constant with one value per triangle, given in closed form:
	\begin{equation*}
		\kappa_T := \frac{(f_b - f_a)\, g_v + (f_a - f_v)\, g_b + (f_v - f_b)\, g_a}{A_T}\,,
	\end{equation*}
	where $A_T = x_a (y_b - y_v) + x_b (y_v - y_a) + x_v (y_a - y_b)$ equals twice the signed area of the triangle $T$~\cite[Eqs.~6, 7]{Kloetzl2022}.
	We assume all triangles to be consistently oriented counterclockwise.
	
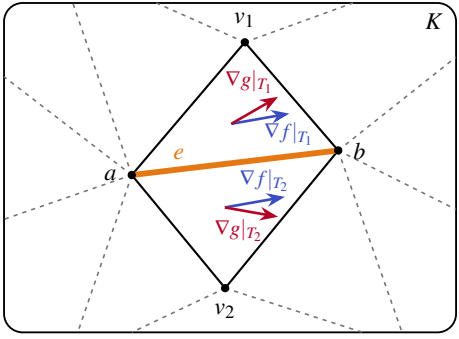
\begin{figure}[bt]
		\centering
		\begin{tikzpicture}[scale=1.3,
			vert/.style={circle,fill=black,inner sep=1.1pt},
			outeredge/.style={black!55,line width=0.6pt,dash pattern=on 1.6pt off 1.9pt},
			grad/.style={-{Stealth[length=2.4mm]},line width=1.0pt}]
			
			\draw[black,line width=0.7pt,rounded corners=7pt] (-1.3,-1.6) rectangle (3.3,1.75);
			\node[anchor=south east] at (3.25,1.4) {$K$};
			
			\coordinate (a)  at (0,0);
			\coordinate (b)  at (2.1,0.25);
			\coordinate (v1) at (1.15,1.35);
			\coordinate (v2) at (0.95,-1.15);
			\draw[outeredge] (a)  -- (-1.3,1.0);
			\draw[outeredge] (a)  -- (-1.3,-0.45);
			\draw[outeredge] (a)  -- (-0.55,-1.6);
			\draw[outeredge] (a)  -- (-0.6,1.75);
			\draw[outeredge] (v1) -- (0.25,1.75);
			\draw[outeredge] (v1) -- (2.25,1.75);
			\draw[outeredge] (b)  -- (3.3,1.2);
			\draw[outeredge] (b)  -- (2.75,-1.6);
			\draw[outeredge] (b)  -- (3.3,-0.35);
			\draw[outeredge] (v2) -- (2.35,-1.6);
			\draw[outeredge] (v2) -- (0.0,-1.6);
			
			\draw[black,line width=0.8pt] (a) -- (v1) -- (b);
			\draw[black,line width=0.8pt] (a) -- (v2) -- (b);
			\draw[jsorange,line width=2.2pt] (a) -- (b);
			\node[jsorange,above=1.5pt] at ($(a)!0.2!(b)+(0.06,-0.02)$) {$e$};
			
			\coordinate (c1) at ($ (a)!0.45!(b)!0.33!(v1) $);
			\draw[grad,cwblue] (c1) -- ++(10:0.6) node[pos=0.5,anchor=north west,inner sep=1pt] {\footnotesize$\nabla f|_{T_1}$};
			\draw[grad,cwred]  (c1) -- ++(30:0.55) node[pos=1,anchor=south east,inner sep=1pt] {\footnotesize$\nabla g|_{T_1}$};
			
			\coordinate (c2) at ($ (a)!0.45!(b)!0.35!(v2) $);
			\draw[grad,cwblue] (c2) -- ++(10:0.62) node[pos=1.15,anchor=south east,inner sep=1pt] {\footnotesize$\nabla f|_{T_2}$};
			\draw[grad,cwred]  (c2) -- ++(-10:0.55) node[pos=0.8,anchor=north east,inner sep=1pt] {\footnotesize$\nabla g|_{T_2}$};
			\foreach \p in {a,b,v1,v2} \node[vert] at (\p) {};
			\node[anchor=east]  at ($(a)+(-0.07,0)$)  {$a$};
			\node[anchor=west]  at ($(b)+(0.07,0)$)   {$b$};
			\node[anchor=south] at ($(v1)+(0,0.07)$)  {$v_1$};
			\node[anchor=north] at ($(v2)+(0,-0.07)$) {$v_2$};
			
		\end{tikzpicture}
		\caption{Critical edge test.
			Two triangles $T_1 = \langle a,b,v_1\rangle$ and $T_2 = \langle a,b,v_2\rangle$ are adjacent to the edge $e=ab$, with constant PL gradients per triangle ($\nabla f$ in blue, $\nabla g$ in red). 
			Since the piecewise constant alignment values $\kappa_{T_1}, \kappa_{T_2}$ change sign across $e$, this edge (orange) is critical and part of the Jacobi set.}
		\label{fig:edge-test}
	\end{figure}

	In the notation of Klötzl~\etal~\cite{Kloetzl2022}, $\kappa_T$ is the linear gradient alignment value $\kappa^{\mathrm{li}}_{v}$ evaluated at the link vertex $v$.
	Comparing gradient alignments of the two triangles $T_1 = \langle a, b, v_1 \rangle$ and $T_2 = \langle a, b, v_2 \rangle$ adjacent to an edge $e=ab$ yields the \emph{critical edge test}~\cite[Eq.~8]{Kloetzl2022}:
	\begin{equation}\label{eq:crit-edge}
		e = ab \in \mathbb{J}_{\mathrm{PL}}(f, g)
		\quad\Longleftrightarrow\quad
		\operatorname{sgn}(\kappa_{T_1}) \neq \operatorname{sgn}(\kappa_{T_2})\,.
	\end{equation}
	Thus, an edge is critical if and only if the piecewise constant alignment field changes sign across it, see the illustration in \cref{fig:edge-test}. 
	
	The collection of critical edges is structured by the \emph{Even Degree Lemma}~\cite{Edelsbrunner2002}: every vertex of $\mathbb{J}_{\mathrm{PL}}$ is incident to an even number of critical edges.
	Therefore, the incident critical edges can be paired at every vertex, without crossings, which unfolds $\mathbb{J}_{\mathrm{PL}}$ into a union of closed curves and mirrors the $1$-manifold structure of the smooth setting~\cite{Edelsbrunner2002}.
	Since the construction only requires a triangulation with vertex values, it applies to arbitrary triangulated $2$-manifolds, including triangulations of scattered data or adaptively refined meshes.
	The restriction of $\mathbb{J}_{\mathrm{PL}}$ to edges of $K$ leads to zig-zag patterns at the resolution of the triangulation~\cite{Kloetzl2022}.
	
	\subsection{Multivariate Gaussians and Quadratic Forms}
	\label{ssec:mvn}
	
	Our new method takes advantage of uncertainty modeling and propagation via multivariate normal distributions.
	An $n$-dimensional multivariate normal random variable $X \sim \mathcal{N}\!(\mu, \Sigma)$ is fully described by two parameters, the mean $\mu \in \mathbb{R}^n$ and the covariance matrix $\Sigma \in \mathbb{R}^{n \times n}$.
	A key property of normal distributions is that they are closed under affine maps with arbitrary target dimensions $k$, i.e., for $A\in \mathbb{R}^{k\times n}$ and $b \in\mathbb{R}^k$
	\begin{equation}\label{eq:affine}
		A X + b \sim \mathcal{N}\!\big(A \mu + b,\; A\Sigma A^\top )\,,
	\end{equation}
	which is the reason for the exact propagation of uncertainty through linear pipelines~\cite[Eq.~2]{Krake2025}.
	
	Even though quadratic transformations of normal distributions are not Gaussian anymore, their moments can be expressed analytically.
	Let $X \sim \mathcal{N}\!(\mu, \Sigma)$ and let $Q$, $Q_1$, and $Q_2$ be symmetric matrices.
	The first two moments of quadratic forms follow from Isserlis' theorem and are classical results in the theory of quadratic forms of Gaussian random variables (see, e.g., Mathai and Provost~\cite{MathaiProvost1992}):
	\begin{align}
		\E\big(X^\top Q\, X\big) &= \tr(Q \Sigma) + \mu^\top Q\, \mu\,, \label{eq:qf-mean}\\
		\Var\big(X^\top Q\, X\big) &= 2\tr(Q \Sigma Q \Sigma) + 4\, \mu^\top Q \Sigma Q\, \mu\,, \label{eq:qf-var}\\
		\Cov\big(X^\top Q_1\, X,\; X^\top Q_2\, X\big) &= 2\tr(Q_1 \Sigma Q_2 \Sigma) + 4\, \mu^\top Q_1 \Sigma Q_2\, \mu\,. \label{eq:qf-cov}
	\end{align}
	A detailed derivation of these moments is given in \autoref{sec:app-qfm}.
	
	\section{Method}\label{sec:method}
	This section discusses the steps that lead to our uncertainty-aware Jacobi set formulation and visualization.
	By modeling the input as multivariate Gaussian~(\cref{ssec:uasf}) and lifting the PL critical edge test to this uncertain input, we obtain a closed-form Jacobi set edge probability~(\cref{ssec:uajs}).
	Then, the edge probabilities are aggregated, serving as a basis for the uncertainty-aware visualization~(\cref{ssec:vis}).
	We close with a discussion of implementation details in~\cref{ssec:impl}.
	
	\subsection{Uncertainty Modeling for Scalar Fields}
	\label{ssec:uasf}
	Given a triangulated compact domain $\mathbb{M}\subset\mathbb{R}^2$ with the triangulation $K$ and
	scalar fields $f,g$ with vertex values $f=(f_1,\ldots, f_N)^\top$ and $g=(g_1,\ldots,g_N)^\top$ at given positions $p_1,\ldots,p_N\in\mathbb{R}^2$, the fields are stacked into one vector and modeled as a multivariate Gaussian distribution as
	\begin{gather*}
		X = \begin{pmatrix}
			f\\g
		\end{pmatrix} \sim \mathcal{N}\!(\mu,\Sigma)\,,\\ \mu = \begin{pmatrix}
			\mu_f\\\mu_g
		\end{pmatrix} \in \mathbb{R}^{2N}, \quad \Sigma = \begin{pmatrix}
			\Sigma_{f\!f} & \Sigma_{fg}\\
			\Sigma_{gf} & \Sigma_{gg}
		\end{pmatrix}\in \mathbb{R}^{2N \times 2N}\,.
	\end{gather*}
	If the fields are not already given as multivariate normal distributions, they should be transformed to such distributions in a first processing step. 
	In the following, we describe two scenarios and ways of acquiring multivariate normal distributions.

	\paragraph{Modeling from Ensemble Data.}
	In many domains, such as weather forecasting or fluid dynamics, the natural origin of uncertain data is simulation runs or repeated measurements.
	Given $E$ ensemble members $x^{(1)}, \ldots, x^{(E)}\in \R^{2N}$ (each member containing both field realizations stacked), the unbiased estimates of mean and covariance of $X$ are:
	\begin{equation*}
		\tilde\mu = \frac{1}{E}\sum_{k=1}^{E}x^{(k)}\,,\quad
		\tilde\Sigma = \frac{1}{E-1}\sum_{k=1}^{E}\big(x^{(k)}-\tilde\mu\big)\big(x^{(k)}-\tilde\mu\big)^\top\,.
	\end{equation*}
	Instead of computing the full $\tilde\Sigma\in\R^{2N\times2N}$, we only store the centered and scaled matrix
	\begin{equation*}
		L=\tfrac{1}{\sqrt{E-1}}\left[x^{(1)}-\tilde\mu, \ldots, x^{(E)}-\tilde{\mu}\right]\in \R^{2N\times E}\,.
	\end{equation*}
	Since typically $E\ll 2N$, this matrix $L$ is smaller than $\tilde{\Sigma}$ and can be used to compute (parts of) $\tilde\Sigma=LL^\top$ on demand.
	
	\paragraph{Modeling via Kernel Functions.}
	Alternatively, the uncertainty is constructed via kernels based on expert knowledge.
	One example is the squared exponential kernel with locally varying uncertainty levels 
	\begin{equation*}
		\big(\Sigma_{f\!f}\big)_{ij} = \sigma_i\, \sigma_j \exp\!\left( - \tfrac{\lVert p_i - p_j \rVert^2}{2\,\ell^2} \right)\,,
	\end{equation*}
	where $\sigma_i\ge0$ is the standard deviation at vertex $i$ and $\ell>0$ the length scale that controls the spatial coherence of the realizations.
	Whereas the kernel specifies the covariance blocks $\Sigma_{f\!f}$ and $\Sigma_{gg}$ directly, a kernel-based construction of the coupling $\Sigma_{fg}$ is not canonical and left to future work.
	As discussed by Krake~\etal~\cite{Krake2025}, Laplacian, periodic, or other kernels can be used as well to model the uncertainty coherence using domain knowledge.
	
\begin{figure}[!tb]
		\centering
		\includegraphics[width=\linewidth]{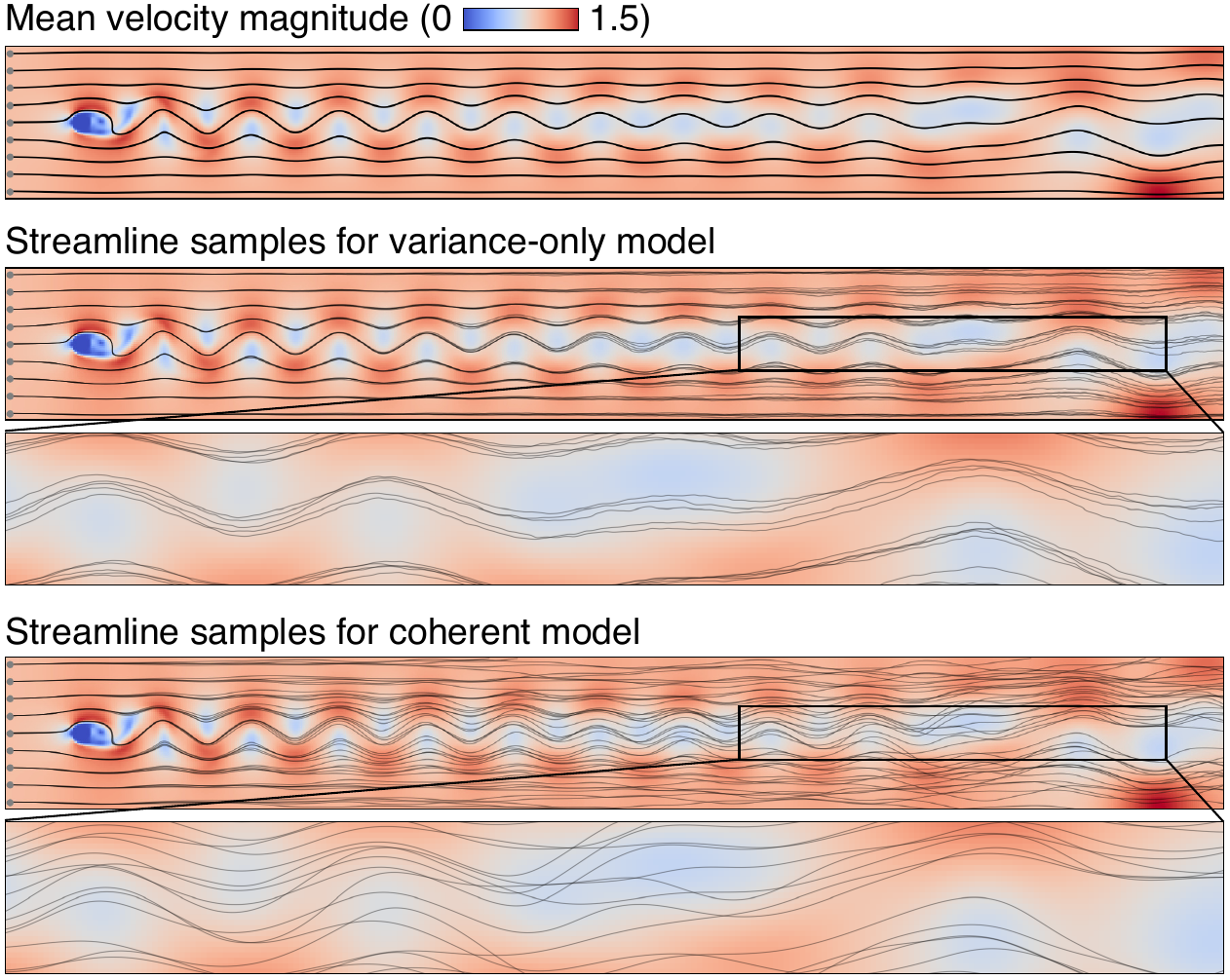}
		\caption{Comparison of spatially coherent uncertainty vs.\ variance-only uncertainty modeling. 
			The two modeled scalar fields $f=u_x$ and $g=u_y$ are the components of the velocity vector field $u=(u_x, u_y)^\top$.
			(Top) Streamlines of the input velocity field, seeded at nine points; the color encodes the velocity magnitude.
			(Middle, bottom) Realizations of $f$ and $g$ drawn from an uncertainty model yield velocity field samples; shown are streamlines of five such samples for the variance-only (middle) and the spatially coherent (bottom) uncertainty model, with corresponding zoom-in areas. 
			In both models, the standard deviation increases from zero at the inflow to $\sigma_{\max}$ at the outflow.}
		\label{fig:unc_model}
	\end{figure}

	\paragraph{Why Covariance Matters: A Vector Field Example.}
	To demonstrate the introduced uncertainty modeling, we apply it to a cylinder flow showing a von Kármán vortex street, a dataset by Günther~\etal~\cite{Guenther17} computed with the Gerris flow solver~\cite{gerrisflowsolver}.
	We compare two different uncertainty models, a spatially coherent one using the squared exponential kernel and a variance-only one without any spatial correlation.
	We model uncertainty on the horizontal and vertical velocity components $f=u_x$ and $g=u_y$ at time step $t=1000$. 
	Put differently, we use the example of a 2D vector field for which we can consider its two vector components as two scalar fields---later for the discussion of the Jacobi set.
	In both models, the standard deviation increases linearly from $\sigma_i=0$ at the inflow to $\sigma_i=\sigma_{\max}$ at the outflow, identically for both components; the coherent model uses the squared exponential kernel, whereas the variance-only model keeps the same $\sigma_i$ but discards all off-diagonal entries.
	The velocity magnitude of the input field, together with the seeded streamlines, is shown in \cref{fig:unc_model}~(top).
	
	To illustrate the different uncertainty models, five pairs of scalar field realizations  (i.e., five vector field realizations) are sampled from each model and streamlines are seeded at identical points.
	Resulting streamlines expose the difference between the variance-only (middle) and spatially coherent (bottom) uncertainty model in~\cref{fig:unc_model}.
	Although both models share the same means and variances and differ only in the off-diagonal structure of $\Sigma$, they induce different flow structures.
	Whereas the coherent model perturbs neighboring vertices consistently into smooth streamlines, the variance-only model introduces independent per-vertex noise and yields jittered streamlines.
	We return to this example in~\cref{sec:eval}, where we discuss the uncertainty model and the resulting uncertainty-aware Jacobi set.
	
	\subsection{Uncertainty-Aware Jacobi Set}
	\label{ssec:uajs}
	We lift the deterministic PL Jacobi set computation to the uncertainty model defined via multivariate normal distributions. 
	This is a three-step process.
	First, the multivariate input is propagated through the PL gradient analytically.
	Second, the probabilistic gradient alignment in individual triangles is derived as quadratic form and the joint gradient alignment values of two adjacent triangles are modeled using second-order moment matching.
	Finally, the per-edge crossing probability of the gradient alignment values is derived.
	
	\paragraph{Gaussian Gradient Field.}
	On each triangle $T=\langle a,b,v\rangle$, the PL gradient of $f$ is constant and given by
	\begin{equation*}
		\nabla f|_T = G_T\, f_T\,, \quad
		G_T = E_T^{-1}\left(\begin{smallmatrix} -1 & 1 & 0 \\ -1 & 0 & 1 \end{smallmatrix}\right) \in \mathbb{R}^{2\times 3}\,, \quad
		E_T = \left(\begin{smallmatrix} (p_b - p_a)^\top \\ (p_v - p_a)^\top \end{smallmatrix}\right)\,,
	\end{equation*}
	where $f_T=(f_a,f_b,f_v)^\top$ collects the vertex values. Thus, $G_T$ depends only on the geometry of $T$, through $E_T$.
	The per-triangle operators are aggregated for both fields $f$ and $g$ into a single global linear operator $\hat{G} \in \R^{4M\times 2N}$, where $M$ is the number of triangles, such that $\hat{G}X$ stacks the four gradient components $(\nabla f|_T, \nabla g|_T)$ of all triangles.
	Since~\cref{eq:affine} applies to any linear map, the stacked gradient vector is again Gaussian:
	\begin{equation*}
		\hat{G}\,X \sim \mathcal{N}\!\big(\hat{G}\mu,\; \hat{G}\, \Sigma\, \hat{G}^\top\big)\,.
	\end{equation*}

	\paragraph{Probabilistic PL Gradient Alignment.}
	Following the critical edge test~\eqref{eq:crit-edge}, an edge $e=ab$ is classified as critical if the gradient alignment changes sign across the two adjacent triangles $T_1$ and $T_2$.
	Restricting the global gradient field $\hat{G}\,X$ to the four gradient components of a single triangle $T$ yields the local Gaussian
	\begin{equation}
		\label{eq:triangle_gaussian}
		Y_T = \hat{G}\,X|_T=\left(\begin{smallmatrix} \partial_x f|_T\\ \partial_y f|_T\\ \partial_x g|_T\\ \partial_y g|_T \end{smallmatrix}\right) \sim \mathcal{N}\!(\mu_T,\Sigma_T)\,,
	\end{equation}
	where $\mu_T$ and $\Sigma_T$ are the corresponding entries from $(\hat{G}\mu,\ \hat{G}\Sigma\hat{G}^\top)$.
	Thus, the gradient alignment in $T$ is obtained as quadratic form via
	\begin{equation*}
		\kappa_T=Y_T^\top QY_T, \quad Q:=\frac{1}{2}\left(\begin{smallmatrix}
			0 & 0 & 0 & 1\\
			0 & 0 & -1 & 0\\
			0 & -1 & 0 & 0\\
			1 & 0 & 0 & 0\\
		\end{smallmatrix}\right)\,.
	\end{equation*}
	The matrix $Q$ is symmetric but indefinite with eigenvalues $\pm\tfrac{1}{2}$, each of multiplicity two.
	Consequently, $\kappa_T$ is an indefinite quadratic form of a Gaussian and therefore follows a generalized chi-squared distribution.
	However, for the edge test, we need the joint distribution of two neighboring gradient alignment values.
	We construct the joint random vector of two adjacent triangles $T_1$ and $T_2$
	\begin{equation*}
		W_e	=\begin{pmatrix}
			Y_{T_1}\\Y_{T_2}
		\end{pmatrix} \sim \mathcal{N}\!(\mu_e,\Sigma_e)\,,
	\end{equation*}
	where $\mu_e$ and $\Sigma_e$ are the corresponding entries of $(\hat{G}\mu,\ \hat{G}\Sigma\hat{G}^\top)$.
	Since $T_1$ and $T_2$ share the edge $ab$, their gradients jointly depend on the shared vertex values, so the cross-triangle block of $\Sigma_e$ is nonzero.
	As a result, this covariance conveys both the shared-vertex coupling and the spatial correlation.
	The two alignment values are now represented as two quadratic forms of this vector:
	\begin{equation*}
		\kappa_{T_1} = W_e^\top Q_1 W_e, \quad \kappa_{T_2} = W_e^\top Q_2 W_e,
	\end{equation*}
	with block matrices $Q_1=\left(\begin{smallmatrix}Q & 0\\ 0 & 0\end{smallmatrix}\right)$ and $Q_2=\left(\begin{smallmatrix}0 & 0\\ 0 & Q\end{smallmatrix}\right)$.
	Both alignments are quadratic forms of the joint Gaussian vector $W_e$, and using the quadratic expressions from \cref{ssec:mvn} for $\mu = \mu_e$ and $\Sigma = \Sigma_e$, their first and second moments are given as 
	\begin{align}
		\mu_i &= \E(\kappa_{T_i}) = \tr(Q_i \Sigma_e) + \mu_e^\top Q_i\, \mu_e\,,\label{eq:kappa-moments-mu}\\
		\sigma_i^2 &= \Var(\kappa_{T_i}) = 2\tr(Q_i \Sigma_e Q_i \Sigma_e) + 4\, \mu_e^\top Q_i \Sigma_e Q_i\, \mu_e\,,\label{eq:kappa-moments-var}\\
		\sigma_{12} &= \Cov(\kappa_{T_1},\kappa_{T_2}) = 2\tr(Q_1 \Sigma_e Q_2 \Sigma_e) + 4\, \mu_e^\top Q_1 \Sigma_e Q_2\, \mu_e\,. \label{eq:kappa-moments-cov}
	\end{align}
	We note for the means, given in~\cref{eq:qf-mean}, that the trace term $\tr(Q\,\Sigma_{T})$ vanishes whenever the two fields are uncorrelated ($\Sigma_{fg}=0$) or symmetrically cross-correlated ($\Sigma_{fg}=\Sigma_{fg}^\top$).
	In this case, $\E(\kappa_T) = \kappa_T(\mu_f, \mu_g)$ coincides with the deterministic alignment of the mean fields.
	While the moments are exact, the joint distribution of $(\kappa_{T_1}, \kappa_{T_2})$ is not Gaussian since each alignment value is a quadratic form.
	Nevertheless, we approximate the distribution using the first- and second-order moments 
	\begin{equation*}
		(\kappa_{T_1},\kappa_{T_2}) \approx \mathcal{N}\!\left(\begin{pmatrix} \mu_1\\\mu_2\end{pmatrix},
		\begin{pmatrix}\sigma_1^2 & \sigma_{12}\\\sigma_{12} & \sigma_2^2\end{pmatrix}\right)\,.
	\end{equation*}
	This second-order moment matching preserves the first two moments exactly but approximates the tails of the joint distribution; we quantify its accuracy against MC references in~\cref{ssec:eval_analytic}.
	
	\paragraph{Edge Probability Computation.}
	The probability that an edge $e$ is critical is given by different signs of the gradient alignment values and can be expressed as
	\begin{equation*}
		p_{e}=\operatorname{P}(\kappa_{T_1}\kappa_{T_2} < 0) = \operatorname{P}(\kappa_{T_1}<0,\,\kappa_{T_2}>0)+\operatorname{P}(\kappa_{T_1}>0,\,\kappa_{T_2}<0)\,.
	\end{equation*}
	Under the approximated Gaussian, standardizing 
	\begin{equation*}
		U=(\kappa_{T_1}-\mu_1)/\sigma_1\,\quad \text{and} \quad V=(\kappa_{T_2}-\mu_2)/\sigma_2
	\end{equation*} yields a bivariate standard normal pair $(U,V)$ with correlation $\rho=\sigma_{12}/(\sigma_1 \sigma_2)$, such that each summand becomes an orthant probability:
	\begin{equation*}
		\operatorname{P}\Big(U<-\tfrac{\mu_1}{\sigma_1},\, V>-\tfrac{\mu_2}{\sigma_2}\Big) = \Phi\Big(-\tfrac{\mu_1}{\sigma_1}\Big) - \Phi_2\Big(-\tfrac{\mu_1}{\sigma_1},-\tfrac{\mu_2}{\sigma_2};\, \rho\Big)\,, 
	\end{equation*}
	with the univariate ($\Phi$) and bivariate ($\Phi_2$) Gaussian cumulative distribution functions.
	Applying this identity to both summands, with $U$ and $V$ exchanged for the second, gives the closed form
	\begin{equation}\label{eq:edge-prob}
		p_{e}=\Phi\left(-\tfrac{\mu_1}{\sigma_1}\right) + \Phi\left(-\tfrac{\mu_2}{\sigma_2}\right) - 2\Phi_2\left(-\tfrac{\mu_1}{\sigma_1},-\tfrac{\mu_2}{\sigma_2};\tfrac{\sigma_{12}}{\sigma_1\sigma_2}\right)\,.
	\end{equation}
	This completes the derivation of the uncertainty-aware critical edge test and we call the probability $p_e$ the \emph{Jacobi set edge probability}.
	
	\subsection{Uncertainty-Aware Jacobi Set Visualization}
	\label{ssec:vis}
	The visualization should communicate three Jacobi-set-specific components: the position of Jacobi set structures, the certainty that these structures occur, and the variability of Jacobi sets under given scalar field realizations.
	The straightforward certainty encoding is color-coding each of the individual edges by its derived Jacobi set edge probability $p_e$ through a continuous colormap.
	In typical applications, the resolution is so high that individual edges are no longer separable.
	We therefore aggregate the edge probabilities to vertices via the \emph{expected Jacobi set (JS) degree} and propose a signed variant that additionally encodes the orientation of the gradient alignment.
	
	Edelsbrunner and Harer~\cite{Edelsbrunner2002} introduced the degree of a vertex $i$ as the number of incident critical edges, $D_i=\sum_{e\ni i}\mathbf{1}\{e \text{ critical}\}$.
	Its expectation follows directly by linearity from the Jacobi set edge probabilities:
	\begin{equation}\label{eq:exp_degr}
		\E(D_i)=\sum_{e\ni i} p_e\,.
	\end{equation}
	By the Even Degree Lemma, $D_i$ is even in every realization. 
	In contrast, \cref{eq:exp_degr} is a probability-weighted average of even integers and, thus, $\E(D_i)$ is in general not even.
	The expected JS degree is exact and bounded by the maximal number of incident edges per vertex, and its interpretation is closely related to the original PL method.
	We recommend binning the encoding to $[0,2]$: whenever the Jacobi set certainly passes a vertex, i.e., two incident edges have $p_e\approx1$, the maximum of the binned degree is reached.
	The converse does not hold and false positives are admitted, as several uncertain incident edges may sum to $2$ or more.
	
	The expected JS degree is nonnegative by construction and therefore discards the sign of the expected alignment $\E(\kappa_T)$, whose sign-change boundary (\cref{eq:crit-edge}) coincides with the Jacobi set of the mean fields when $\E(\kappa_T)=\kappa_T(\mu_f,\mu_g)$ holds, e.g., for $\Sigma_{fg}=0$. 
	The \emph{signed expected JS degree}
	\begin{equation}\label{eq:signed_degr}
		\tilde{D}_i = s_i\, \E(D_i)\,, \qquad s_i = \operatorname{sgn}\Big(\sum_{T\ni i} \E(\kappa_T)\Big) \in \{-1,+1\}\,,
	\end{equation}
	retains this orientation.
	We use the unsigned expected JS degree by default and the signed variant if the alignment sign carries meaning.
	
	These vertex-based aggregations enable the following layered visual representation, tailored to the characteristics of Jacobi sets.
	The \emph{Jacobi set of the mean fields} $(\mu_f, \mu_g)$, in short the mean Jacobi set, encodes the position and explicit line geometry that remains readable even where the heatmap values are close to zero.
	To be concrete, the mean Jacobi set is not an expected or most probable Jacobi set but the result that a deterministic Jacobi set of the mean fields would show and thereby also serves as reference.
	For the deterministic Jacobi set, different edge-based methods can be used, such as the PL or local bilinear computation of Jacobi sets with reduced connectivity~\cite{Kloetzl2022, Kloetzl2022a}.
	We extract the mean Jacobi set throughout this work via the PL method.
	The certainty is encoded by the binned expected JS degree as a sequential blue heatmap, such that dark, saturated bands confirm mean Jacobi set structures, whereas light, desaturated regions indicate uncertainty in the Jacobi set position and occurrence and thereby separate persistent structures from numerically sensitive ones visually. 
	For the signed variant, mapping $\tilde{D}_i$ with a diverging blue--red colormap centered at zero colors the two sides of the probabilistic band by their alignment values: red for $\E(\kappa)\ge0$ and blue for $\E(\kappa)<0$, separated along the interface by the sign change of $\E(\kappa)$.
	The realization-level variability can be shown by drawing individual samples from the input model $\mathcal{N}\!(\mu,\Sigma)$ and computing the respective Jacobi set of each sample.
	As with the mean Jacobi set, the same edge-based computation methods can be used, and each derived Jacobi set sample is then overlaid in a distinct color; the samples are drawn independently and illustrate the line geometry of single realizations.
	\Cref{fig:teaser} (right) shows the complete composition of all three layers for a weather ensemble.
	
	\begin{algorithm}[b]
		\caption{Uncertainty-aware Jacobi set.}
		\label{alg:uajs}
		\textbf{Input:} triangulation $K$ with vertex positions $p_i$; means $\mu_f, \mu_g$; covariance access to $\Sigma$.\\
		\textbf{Output:} edge probabilities $p_e$; vertex aggregates $\E(D_i)$ and $\tilde{D}_i$.
		\begin{algorithmic}[1]
			\State \textbf{for each} triangle $T$ \textbf{do}
			\State \hspace{3pt}construct $G_T$ and compute $\mu_T, \Sigma_T$ of \cref{eq:triangle_gaussian}
			\State \hspace{3pt}evaluate and cache $\E(\kappa_T)$, $\Var(\kappa_T)$ via \cref{eq:qf-mean,eq:qf-var}
			\State \textbf{end for}
			\State \textbf{for each} edge $e$ with adjacent $T_1, T_2$ \textbf{do}
			\State \hspace{3pt}gather the joint moments $(\mu_e, \Sigma_e)$ of $W_e = (Y_{T_1}, Y_{T_2})$
			\State \hspace{3pt}access $\mu_1, \mu_2, \sigma_1^2, \sigma_2^2$ and compute $\sigma_{12}$ via \cref{eq:kappa-moments-mu,eq:kappa-moments-var,eq:kappa-moments-cov}
			\State \hspace{3pt}compute $p_e$ via \cref{eq:edge-prob}
			\State \textbf{end for}
			\State aggregate $\E(D_i)$ and $\tilde{D}_i$ via \cref{eq:exp_degr,eq:signed_degr} for each vertex $i$
			\State \Return $p_e$, $\E(D_i)$, $\tilde{D}_i$
		\end{algorithmic}
	\end{algorithm}
	
	\subsection{Implementation}\label{ssec:impl}
	
	\Cref{alg:uajs} summarizes the analytic pipeline.
	Even though the input model is a 2$N$-dimensional Gaussian (with $N$ the number of vertices of $K$), the algorithm never computes the full $\Sigma\in\R^{2N\times 2N}$. 
	Each triangle only requires the $6\times 6$ block of $\Sigma$ over the values of both fields at its three vertices $\{a,b,v\}$, from which the $4\times 4$ gradient covariance $\Sigma_T$~\eqref{eq:triangle_gaussian} is computed. 
	Each interior edge $e=ab$ requires the $8\times 8$ block over the four vertices $\{a,b,v_1,v_2\}$ of its two adjacent triangles $T_1$ and $T_2$ to compute the covariance $\Sigma_e$ of $W_e$.
	Both blocks are built either from the corresponding rows of $L$ in the ensemble case or by evaluating the kernel.
	
	The algorithm is divided into two loops.
	First (in lines 1--4), the PL gradient operator is constructed for every triangle and the alignment mean $\E(\kappa_T)$ and variance $\Var(\kappa_T)$ are cached.
	In the second loop, the joint distribution $(\kappa_{T_1}, \kappa_{T_2})$ is approximated for each edge by using the cached moments of the two triangles $T_1$ and $T_2$ and computing the cross-covariance $\sigma_{12}$ via~\cref{eq:qf-cov}, such that the Jacobi set edge probability can be computed in line 8.
	Finally, the expected JS degree and its signed variant are aggregated for each vertex in line 10.
	These local computations are a practical advantage over the MC baseline that draws full field realizations from the Gaussian input and extracts one Jacobi set per realization.
	
	\section{Evaluation}\label{sec:eval}
	We evaluate the method in three different settings.
	First, we validate the analytic edge probabilities against an MC reference on the analytic dataset of Klötzl~\etal~\cite{Kloetzl2022, Kloetzl2022a} in~\cref{ssec:eval_analytic}.
	Second, we apply different uncertainty models to a given fluid flow simulation dataset and compare the resulting uncertainty-aware Jacobi set~(\cref{ssec:eval_kernels}).
	Finally, the method is applied to real-world weather ensemble data~(\cref{ssec:eval_real_world}).

	\subsection{Evaluation for Analytic Dataset}
	\label{ssec:eval_analytic}
	
	We start the evaluation by comparing the analytic Jacobi set edge probability against an MC baseline in a fully analytic example.
	All methods were implemented in Python and the evaluation was performed in Jupyter notebooks.
	All computations were executed on a MacBook Pro with an Apple M4 Pro CPU and 24\,GB of RAM.
	The code is publicly available.\footnote{\href{https://github.com/VisDan93/uncertainty-aware-jacobi-set}{\nolinkurl{github.com/VisDan93/uncertainty-aware-jacobi-set}}}
	
\begin{figure}[!tb]
		\centering
		\includegraphics[width=\linewidth]{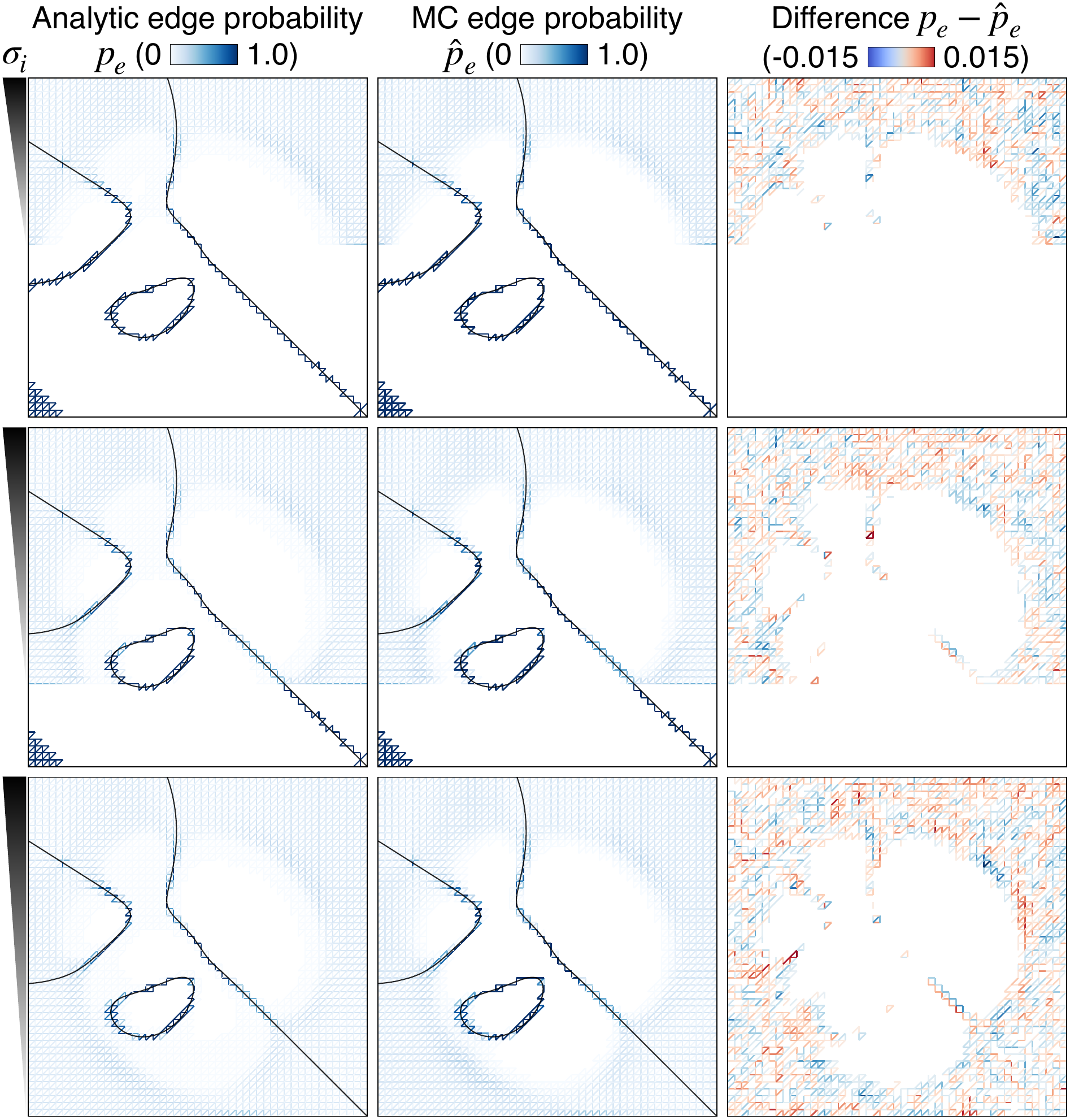}
		\caption{Comparison of analytic Jacobi set edge probability $p_e$ (first column) and MC estimate $\hat{p}_e$ (second column) for the analytic field of Klötzl~\etal~\cite{Kloetzl2022}, with gradually increasing uncertainty in the upper half, the upper three quarters, and the entire domain (from top to bottom).
			The third column illustrates the per-edge difference of the two edge probabilities from the analytic and the MC computation.}
		\label{fig:mc_comparison}
	\end{figure}

	We use the analytic dataset introduced by Klötzl~\etal, given by two scalar fields on $[-1,1]^2$, each defined as sum of three multivariate normal distributions.
	Since both fields are smooth and given in closed form, the gradient alignment field and Jacobi set are known analytically and thereby represent a useful test case for the method and implementation.
	The fields are sampled with a resolution of $50 \times 50$.
	The input uncertainty is modeled using the squared exponential kernel described in~\cref{ssec:uasf} with length scale $\ell=0.15$ and standard deviation $\sigma_i=0$ up to a specific $y$-value. 
	Starting at this value, $\sigma_i$ is increasing linearly, reaching $1\,\%$ of the scalar fields' standard deviation at $y=1$.
	This results in three different configurations (uncertainty ramp starting at $y=\{0.0, -0.5, -1.0\}$).
	To keep the model simple, there is no covariance between the fields modeled ($\Sigma_{fg}=0$).
	\Cref{fig:mc_comparison} provides visualizations of the dataset, with the three configurations shown in the rows ordered from top to bottom.
	
	\paragraph{Monte Carlo Reference.}
	The MC comparison draws $N_{\mathrm{MC}}$ realizations $x^{(k)}\sim\mathcal{N}\!(\mu,\Sigma)$ from the input uncertainty model and applies the deterministic critical edge test~\eqref{eq:crit-edge} to each realization.
	The per-edge crossing frequency is averaged over all $N_{\mathrm{MC}}$ realizations:
	\begin{equation*}
		\hat{p}_e = \frac{1}{N_{\mathrm{MC}}}\sum_{k=1}^{N_{\mathrm{MC}}} \mathbf{1}\big\{\operatorname{sgn}\kappa_{T_1}^{(k)} \neq \operatorname{sgn}\kappa_{T_2}^{(k)}\big\}\,,
	\end{equation*}
	which converges to the edge probability for $N_{\mathrm{MC}}\to \infty$.
	Our analytic $p_e$ approximates this frequency without sampling, using the second-order moment matching approximation in~\cref{ssec:uajs}.
	
	\paragraph{Visual and Numerical Comparison.}
	Each MC reference uses $N_{\mathrm{MC}}=10{,}000$ realizations, and we draw $100$ independent references per configuration.
	\Cref{fig:mc_comparison} arranges the comparison for the three described uncertainty models as a $3\times 3$ matrix, one row per uncertainty configuration consisting of columns illustrating our analytic $p_e$, the MC $\hat{p}_e$, and the difference per edge. 
	The first two columns are visually identical and the uncertain region around the Jacobi set widens equally in both methods. In the certain areas, the probabilities are binary, $p_e\in\{0,1\}$, and reproduce the PL Jacobi set.
	Compared to the analytic Jacobi set~\cite[Fig.~7]{Kloetzl2022}, the discrete views show degenerate areas on the bottom left and right parts of the domain. 
	This already hints at unstable areas that are directly affected by the introduced uncertainty. 
	The difference column shows unstructured errors with magnitudes below~0.015.
	For a quantitative comparison, we measure the relative mean absolute difference $\lVert p_e - \hat{p}_e \rVert_1 / \lVert\hat{p}_e\rVert_1$ and average it over the 100 references.
	Our analytic solution deviates from the references by $1.2\,\%$ to $1.8\,\%$ across the three configurations, whereas independent references deviate from each other by $1.6\,\%$ to $2.5\,\%$. 
	The deviation of our analytic solution therefore stays below the spread of the references themselves.
	
	By inspecting the visualizations of the three variants of uncertainty models, we can recognize that the area with radius larger $\approx0.7$ around the center is affected by increasing uncertainty, resulting in wide and low-valued ($p_e<0.3$) uncertain Jacobi set areas. In contrast, the center shows robust Jacobi set structures, e.g., the Jacobi set structure running from the top to the bottom right remains stable under increasing uncertainty, whereas in the second and third rows, the bottom-right region shows variability between neighboring Jacobi set edges around the analytic solution.
	
	On the $50\times50$ grid, the analytic Jacobi set computation takes 0.3\,s for the configuration with uncertainty over the whole domain, whereas one MC reference requires 1.7\,s, including the drawing of the field samples.
	Both implementations are parallelizable either over independent triangles and edges with cache-coherent memory access (ours) or over realizations (MC).
	
	\begin{figure}[!tb]
		\centering
		\includegraphics[width=\linewidth]{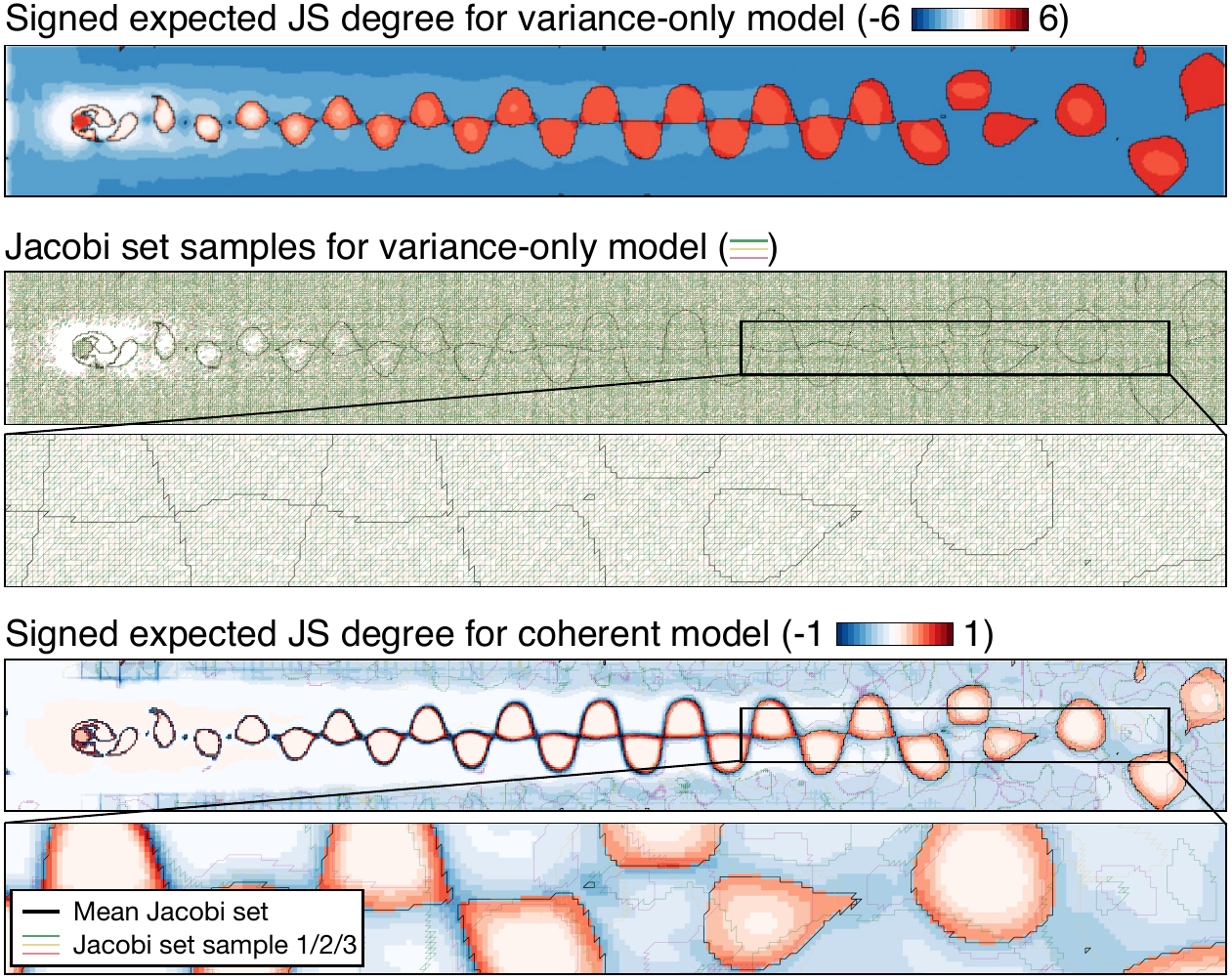}
		\caption{Uncertainty-aware Jacobi set for the variance-only (top and middle) and the spatially coherent uncertainty model (bottom row), including zoom-in areas.
			Variance increases from left to right. 
			The signed expected JS degree encodes the Jacobi set density via saturation and the alignment sign via hue (red: rotation-dominated; blue: strain-dominated).
			Jacobi set samples of three field realizations are visualized as colored lines. 
			Due to the substantial sample overlap in the variance-only case, the signed expected JS degree and samples are shown separately.
			For the coherent model, the signed expected JS degree, samples, and mean are shown in one view.
		}
		\label{fig:eval_karman}
	\end{figure}
	\subsection{User-Defined Kernels for a Flow Simulation}
	\label{ssec:eval_kernels}
	
	We continue with the vector field example from~\cref{ssec:uasf}, where we applied two different uncertainty models to the cylinder flow with the von Kármán vortex street dataset by Günther~\etal~\cite{Guenther17}, and examine the derived uncertainty-aware Jacobi set.
	The uncertainty is modeled onto the $x$- and $y$-velocity components at time step $t=1000$.
	The standard deviation increases linearly from zero at the inflow to $\sigma_{\max}=0.15$ at the outflow (about 15\,\% of the mean flow speed), with correlation length $\ell = 12$ grid cells for the coherent model.
	
	In the streamline comparison in~\cref{fig:unc_model}, the locally independent noise averages out along the integration and the uncertainty is barely noticeable, except for the smaller scale jittering of the streamlines, whereas the coherent model displaces whole streamlines and exposes a plausible spread of the flow.
	
	For this example, the gradient alignment has a direct physical interpretation.
	For the component pair $(f,g)=(u_x,u_y)$, the alignment value equals the Jacobian determinant of the velocity field, $\kappa = \det (\nabla u)$, which is positive in rotation-dominated regions and negative in strain-dominated areas.
	The Jacobi set is the boundary of these flow regimes and the signed expected JS degree $\tilde{D}_i$~\eqref{eq:signed_degr} colors the two areas accordingly in~\cref{fig:eval_karman}.
	The PL gradient takes differences of neighboring vertex values, so the independent per-vertex noise of the variance-only model amplifies in the gradient. 
	Since the alignment $\kappa_T$ is built from these gradients, the noise dominates it and the Jacobi set results in an unstructured noise across the domain, except for the region near the cylinder. 
	This can be noted when examining the three samples in the variance-only uncertainty model (\cref{fig:eval_karman}, middle) that almost build a space-filling Jacobi set across the entire domain.
	The signed expected JS degree of the variance-only model verifies this observation (\cref{fig:eval_karman}, top): except near the low-uncertainty inflow, the expected JS degree saturates the entire domain and the red--blue colormap shows the alignment regimes of the mean flow.
	The coherent model instead keeps neighboring vertex values and thereby also gradients correlated such that the alignment stays meaningful and the visualized samples result in interpretable Jacobi set lines that only move away from the mean when the uncertainty increases (\cref{fig:eval_karman}, bottom).
	Our signed expected JS degree confirms this observation by tracing the mean Jacobi set closely in the low-uncertainty area and widening as the uncertainty grows (\cref{fig:eval_karman}, bottom).
	Furthermore, this example illustrates that adding individual samples to the visualization benefits the comprehension of the underlying model since individual samples can also deviate from the expected JS degree heatmap, e.g., the samples leave the probable Jacobi set area with growing uncertainty more frequently.
	
	This use case showcases uncertainty modeling for scalar fields and the benefit of a carefully designed (spatially coherent) uncertainty model.
	Especially, showing only the mean Jacobi set is often misleading and if one is interested in the Jacobi distribution, it is beneficial to include individual Jacobi set samples into the analysis
	(see the bottom two plots in~\cref{fig:eval_karman}).
	This comparison also explains why the streamlines in~\cref{fig:unc_model} and the Jacobi sets in~\cref{fig:eval_karman} react differently to the same variance-only model.
	Streamline integration averages the noise along the integration path such that independent per-vertex noise largely cancels.
	On the other hand, the PL gradient differences neighboring vertex values such that the same noise is amplified.
	A variance-only model understates the uncertainty of integral curves and overstates the uncertainty of methods that depend on derivatives, such as the Jacobi set, whereas the spatially coherent model shows a more consistent behavior for both.
	
	Note that, in this setting, the two scalar fields are not coupled in the uncertainty model and it holds $\Sigma_{fg}=0$.
	This cross-field coupling is another modeling choice left to future work.
	
	\begin{figure*}[!tb]
		\centering
		\includegraphics[width=\linewidth]{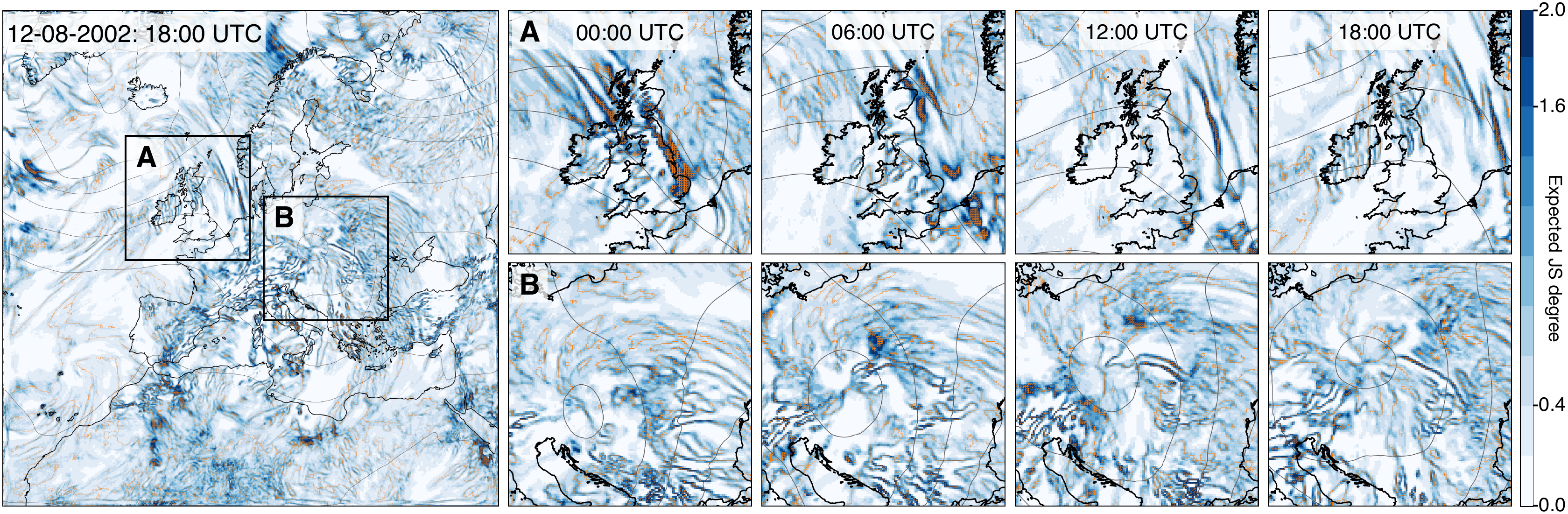}
		\includegraphics[width=\linewidth]{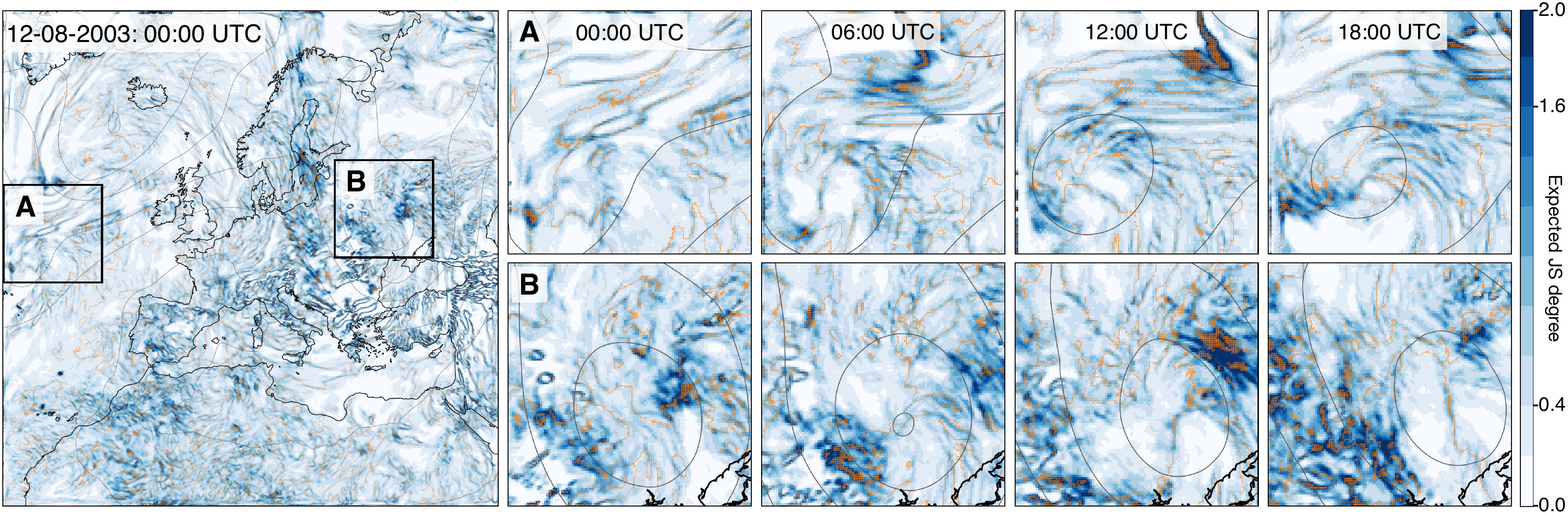}
		\caption{Uncertainty-aware Jacobi set for the CERRA-EDA weather ensemble (\emph{E} = 10 members) of temperature and geopotential at 500\,hPa, for the 2002 European floods (top block, 12~August 2002) and the 2003 European heat wave (bottom block, 12~August 2003).
			Each block shows a full-domain overview at the indicated reference time with the zoom regions A and B marked; both regions are tracked over the four analysis times 00:00, 06:00, 12:00, and 18:00\,UTC.
			The visualization consists of the mean Jacobi set (orange), the expected JS degree $\E(D_i)$ as a blue heatmap, and coastlines with mean geopotential isolines for context. }
		\label{fig:eval_cerra}
	\end{figure*}
	
	\subsection{Evaluation of Real-World Weather Ensemble Data}
	\label{ssec:eval_real_world}
	
	The previous use case modeled uncertainty via user-defined kernels since the uncertainty was not directly given in the data.
	We now use data whose uncertainty is directly given via ensemble members.
	The Copernicus European Regional Reanalysis (CERRA)~\cite{Ridal2024} reconstructs the atmospheric state over Europe by assimilating observations into a numerical weather prediction model.
	Its analysis uncertainty is quantified by the ensemble data assimilation system CERRA-EDA that consists of $E=10$ members on a Lambert conformal conic grid of $565\times 565$ vertices with 11\,km spacing every 6 hours.
	As scalar field inputs for the Jacobi set computation, we use the temperature (in K) and geopotential (in m$^2$s$^{-2}$) at the 500\,hPa pressure level (about 5--6\,km above the ground).
	
	Using these two fields for a Jacobi set computation is meteorologically motivated.
	At this pressure level, the wind blows approximately along isolines of the geopotential (geostrophic wind).
	When isolines of the temperature cross the geopotential isolines, the wind moves warmer or colder air into a region, and when the isolines run in parallel, no temperature is moved.
	Thus, the gradient alignment value of temperature and geopotential is proportional to the temperature transport by the geostrophic wind.
	As a result, the Jacobi set consists of boundaries between regions that the flow warms and regions that it cools.
	The visualized uncertainty then describes how well these boundaries hold throughout the ensembles.
	
	At each analysis time, the members are stacked into $x^{(k)}\in\R^{2N}$ and the ensemble mean and covariance are estimated.
	The estimated covariance contains a nonzero cross-field covariance $\Sigma_{fg}$ between temperature and geopotential, as expected, because the two fields are physically coupled.
	We examine two distinct extreme weather events in the European region at the times 00:00, 06:00, 12:00, and 18:00\,UTC.
	The first weather event is the 2002 European floods, where slowly moving humid air was pushed against the central European mountain ranges, resulting in long-enduring rainfalls that peaked on 12~August 2002~\cite{Ulbrich2003}.
	The second date, 12~August 2003, is during the peak of the European heat wave that built up under an enduring omega blocking over central Europe~\cite{Black2004}.
	\Cref{fig:eval_cerra} shows the resulting uncertainty-aware Jacobi set visualized via the Jacobi set of the mean fields (in orange), the expected JS degree $\E(D_i)$ from~\cref{eq:exp_degr} binned on $[0,2]$ with a blue heatmap, and the coastlines and mean geopotential isolines for context.
	For each date, an overview of the full domain is accompanied by two zoom regions A and B that are tracked over the four analysis times.
	
	\paragraph{General Observations.}
	For the given grid resolution, the mean Jacobi set spreads over the entire domain because it even captures very small-scale gradient alignments.
	When explored separately, the viewer cannot distinguish strongly separating Jacobi set structures from numerically sensitive ones.
	The expected JS degree adds this missing information.
	Narrow dark bands with $\E(D_i)$ close to 2 or higher reveal areas where a Jacobi set line passes for almost every ensemble member.
	On the other side, pale areas locate sections where the Jacobi sets have a higher spatial variability within the ensemble.
	In addition, larger-scale structures often persist over the four analysis times, whereas Jacobi sets at grid scale vary.
	
	\paragraph{12 August 2002: European Floods.}
	\Cref{fig:eval_cerra} (top) visualizes data for the European floods.
	In zoom region A, we can study an area with high expected JS degree over a dense mean Jacobi set cluster along the east coast of Great Britain.
	Over the day, this cluster decays and subdivides; from 12:00\,UTC one, and at 18:00\,UTC two almost parallel structures remain.
	The mean Jacobi set verifies these areas of gradient alignments.
	Using the uncertain Jacobi set lets us follow the movement of robust Jacobi set features more easily.
	
	Zoom region B contains the low-pressure area that led to the flood rainfall, recognizable by the nearly circular isolines that move northeast over the day.
	Between 00:00\,UTC and 12:00\,UTC, high-probability Jacobi set structures appear in the south-west of the low-pressure region, which corresponds to the areas where the heaviest rainfall was recorded~\cite{Ulbrich2003}.
	The rotation direction of the low (anti-clockwise) is reflected in this region and can be depicted by the circular arcs in the east of the low.
	
	\paragraph{12 August 2003: European Heat Wave.}
	The heat wave produces a different environment, see \cref{fig:eval_cerra} (bottom).
	First, compared to the flood situation, the zoomed-out view contains larger areas with positive Jacobi set probability and especially over central Europe and the region of the Alps, the uncertainty-aware Jacobi set is much more washed-out.
	This is reinforced by the widely spaced isolines from Iberia to central Europe, which indicate weak air movement.
	In this stable weather situation, the gradients in both fields are small, the sign of the gradient alignment becomes unstable, and larger areas with low Jacobi set probabilities evolve.
	Looking only at the mean Jacobi set would thereby be misleading.
	In zoom region A, a rotating low forms during the day with a high expected JS degree area building up in the north-east of the low and disappearing again.
	In region~B, which is closer to the stable heat wave region, no such strong rotational effects can be observed in the Jacobi set.
	This illustrates a property of Jacobi sets, i.e., that they pass through critical points of the individual scalar fields.

	\section{Discussion and Conclusion}
	We presented an analytic uncertainty-aware Jacobi set computation.
	Based on an uncertainty modeling of joint scalar fields via multivariate normal distributions, the uncertainty is propagated through the PL gradient computation, and the critical edge test is lifted to a closed-form edge probability.
	The evaluation showed a close agreement with MC references and demonstrated the versatility of the method for user-defined kernel models and real-world ensemble data.
	For visual analysis, the derived encodings separate robust Jacobi set structures from numerically sensitive ones and thereby provide a more honest representation that motivated our approach.
	Even though the Gaussian uncertainty model introduces high-dimensional covariance matrices for high spatial resolutions, our method scales to large triangulations without forming the full covariance matrix.
	
	Our multivariate uncertainty model introduces flexibility in modeling and enables an exact propagation of means and covariances.
	Moreover, we emphasized that spatially coherent Gaussians are a good fit for smooth physical fields where the continuity is reflected by local correlation.
	However, this model also introduces limitations if explicitly non-Gaussian input uncertainty is given.
	Furthermore, the second-order moment matching for computing the Jacobi set edge probability introduces approximation errors, which showed to be small in our comparison against MC references.
	
	Beyond two dimensions, the edge-based identification of Jacobi sets extends to triangulated 3-manifolds~\cite{Edelsbrunner2002}, and the PL gradients remain linear in the vertex values.
	However, the link of an edge then contains several vertices, so the pairwise sign-change probability needs to be generalized to higher-dimensional orthant probabilities, which we leave to future work.
	Another branch for future work could cover a detailed analysis of the proposed uncertainty model for scalar fields, i.e., kernel-based constructions of the cross-field covariance, the reliability of covariance estimates from few ensemble members, and extensions beyond Gaussians to capture multimodal uncertainty.

	\section*{AI Use Statement}
	Large language models were employed as assistive tools in the preparation of this work.
	Claude Opus 4.8 and Claude Fable 5 supported the writing process through proofreading, language editing, and verification of notation, references, and mathematical derivations; the scientific ideas, the method, and all results originate from the authors.
	Claude Sonnet 4.6 assisted in writing and debugging prototype code.
	All figures were produced by the authors' own implementation.
	The authors reviewed and verified all AI-assisted content and take full responsibility for the manuscript.
	
	\appendix
	\section*{APPENDIX}
	\section{Moments of Quadratic Forms of Multivariate Normal Distributions}\label{sec:app-qfm}
	
	Here, we provide derivations and more details for the equations summarized in \cref{ssec:mvn}.
	
	Let $X \sim \mathcal{N}\!(\mu, \Sigma)$ with $\mu \in \R^n$ and $\Sigma \in \R^{n \times n}$ be multivariate normal and $Q, Q_1, Q_2 \in \R^{n \times n}$ symmetric.
	We write $X = \mu + Z$ with $Z := X - \mu \sim \mathcal{N}\!(0, \Sigma)$, so that $\E(Z_i Z_j) = \Sigma_{ij}$.
	Isserlis' theorem expresses higher moments of a centered Gaussian vector through its pairwise covariances~\cite{Isserlis1918}.
	We use two of its consequences.
	First, products with an odd number of factors have vanishing expectation, e.g., $\E(Z_i)=\E(Z_iZ_jZ_k)=0$.
	Second, the expectation of an even number of factors is obtained by summing the covariance products over all perfect pairings, e.g., for a four-factor moment, it holds
	\begin{equation*}
		\E(Z_i Z_j Z_k Z_l) = \Sigma_{ij} \Sigma_{kl} + \Sigma_{ik} \Sigma_{jl} + \Sigma_{il} \Sigma_{jk}\,.
	\end{equation*}
	We now derive the mean~(see \cref{eq:qf-mean}), variance~(see \cref{eq:qf-var}), and covariance~(see \cref{eq:qf-cov}) one by one.
	\paragraph{Mean.}
	Expanding the quadratic form using $X = Z+\mu$ and $Q = Q^\top$ yields
	\begin{equation*}
		X^\top Q\, X = \underbrace{\mu^\top Q\, \mu}_{A} + \underbrace{2\, \mu^\top Q\, Z}_{B} + \underbrace{Z^\top Q\, Z}_{C}\,,
	\end{equation*}
	with deterministic $A$, linear $B$ (in $Z$) with $\E(B)=0$, and 
	\begin{equation*}
		\E(Z^\top Q\, Z) = \sum_{i,j} Q_{ij}\, \E(Z_i Z_j) = \sum_{i,j} Q_{ij} \Sigma_{ji} = \tr(Q \Sigma)\,.
	\end{equation*}
	Thereby, the expectation of $X^\top Q\, X$ is~\eqref{eq:qf-mean}.
	
	\paragraph{Variance.}
	Since $A$ is constant, 
	\begin{equation*}
		\Var(X^\top Q\, X) = \Var(B + C) = \Var(B) + 2\Cov(B, C) + \Var(C)\,.
	\end{equation*}
	Since $\E(B)=0$, we have $\Cov(B,C)=\E(BC)-\E(B)\E(C)=\E(BC)$ and with $u:=Q\mu$, this cross-term becomes zero,
	\begin{equation*}
		\E(B C) = 2 \sum_{i,j,k} u_i\, Q_{jk}\, \E(Z_i Z_j Z_k) = 0\,.
	\end{equation*}
	Writing $B=(2Q\mu)^{\top}Z$ as a scalar linear form and applying $\Var(a^\top Z) = a^\top \Sigma a$ gives
	\begin{equation*}
		\Var(B) = (2 Q \mu)^\top \Sigma\, (2 Q \mu) = 4\, \mu^\top Q \Sigma Q\, \mu\,.
	\end{equation*}
	For $\Var(Z^\top Q Z)$, the four-factor Isserlis' moment is applied:
	\begin{align*}
		\E\big((Z^\top Q\, Z)^2\big) &= \sum_{i,j,k,l} Q_{ij} Q_{kl}\, \E(Z_i Z_j Z_k Z_l) \\
		&= \sum_{i,j,k,l} Q_{ij} Q_{kl} \big(\Sigma_{ij} \Sigma_{kl} + \Sigma_{ik} \Sigma_{jl} + \Sigma_{il} \Sigma_{jk}\big)\,.
	\end{align*}
	By using $\Sigma=\Sigma^\top$ and $Q=Q^\top$, the three terms can be written as the following trace expressions: 
	\begin{gather*}
		\sum_{i,j,k,l} Q_{ij} Q_{kl}\, \Sigma_{ij} \Sigma_{kl} = \big(\!\tr(Q \Sigma)\big)^2\,, \quad 
		\sum_{i,j,k,l} Q_{ij} Q_{kl}\, \Sigma_{ik} \Sigma_{jl} = \tr(Q \Sigma Q \Sigma)\,, \\
		\sum_{i,j,k,l} Q_{ij} Q_{kl}\, \Sigma_{il} \Sigma_{jk} = \tr(Q \Sigma Q \Sigma)\,.  
	\end{gather*}
	This leads to $\E((Z^\top Q Z)^2)=\tr(Q\Sigma)^2 + 2\tr(Q\Sigma Q\Sigma)$.
	Subtracting $(\E(Z^\top Q Z))^2$ cancels the first term and results in:
	\begin{equation*}
		\Var(C) = \Var(Z^\top Q\, Z) = 2\tr(Q\Sigma Q\Sigma)\,.
	\end{equation*}
	Adding $\Var(B)$ and $\Var(C)$ establishes \cref{eq:qf-var}.
	
	\paragraph{Covariance.}
	The covariance uses the same arguments known from the variance computation in parallel.
	Let $B_i:=2\mu^\top Q_iZ$ and $C_i:=Z^\top Q_iZ$ for $i\in \{1,2\}$. Dropping the constant parts yields
	\begin{align*}
		\Cov\big(X^\top Q_1\, X,\, X^\top Q_2\, X\big) ={}& \Cov(B_1, B_2) + \Cov(B_1, C_2) \notag\\
		&+ \Cov(C_1, B_2) + \Cov(C_1, C_2)\,.
	\end{align*}
	As above, the two mixed terms $\Cov(B_1,C_2)$ and $\Cov(C_1,B_2)$ are $0$. The linear part is again expressed as scalar linear forms and results in:
	\begin{equation*}
		\Cov(B_1, B_2) = (2 Q_1 \mu)^\top \Sigma\, (2 Q_2 \mu) = 4\, \mu^\top Q_1 \Sigma Q_2\, \mu\,.
	\end{equation*}
	The quadratic part leads to the analog trace formulations as above, with $Q_1$ and $Q_2$ in place of $Q$, respectively. Thus, we obtain:
	\begin{equation*}
		\E(C_1 C_2) = \tr(Q_1 \Sigma) \tr(Q_2 \Sigma) + 2\tr(Q_1 \Sigma Q_2 \Sigma)\,.
	\end{equation*}
	Since $\E(C_i)=\tr(Q_i\Sigma)$, the first term cancels again and we get 
	\begin{equation*}
		\Cov(C_1, C_2) = 2\tr(Q_1 \Sigma Q_2 \Sigma)\,.
	\end{equation*}
	Adding $\Cov(B_1, B_2)$ and $\Cov(C_1, C_2)$ leads to~\cref{eq:qf-cov}.
	
	\acknowledgments{\noindent 
		This work was supported by the Deutsche Forschungsgemeinschaft (DFG, German Research Foundation) under Germany’s Excellence Strategy -- EXC 2120/2 -- 390831618 and Project-ID 251654672 -- TRR 161.
		The authors thank Tim Krake for feedback on the mathematical formalization of the paper.
	}
	
	\bibliographystyle{abbrv-doi-hyperref}
	\bibliography{bib}

\end{document}